\documentclass{spie}
\usepackage{epsfig,emp,latexsym}

\newcommand{\nix}[1]{}
\newcommand{\mat}[4]{\left(\begin{array}{rr}#1& #2\\ #3 & #4\end{array}\right)}
\newcommand{\cmat}[4]{\left(\begin{array}{cc}#1& #2\\ #3 & #4\end{array}\right)}

\newcommand{\onemat}{\mathbf{1}}
\newcommand{\zeromat}{\mathbf{0}}
\newcommand{\ket}[1]{\left|#1\right\rangle}
\newcommand{\qed}{\mbox{$\Box$}}
\newcommand{\C}{\mathbf{C}}
\newcommand{\diag}{\mbox{diag}}

\newcommand{\adots}{%
\raisebox{0ex}{\mbox{$\cdot$}}%
\,\raisebox{0.75ex}{\mbox{$\cdot$}}%
\,\raisebox{1.5ex}{\mbox{$\cdot$}}%
}
\newcommand{\bdots}{%
\raisebox{1.5ex}{\mbox{$\cdot$}}%
\,\raisebox{0.75ex}{\mbox{$\cdot$}}%
\,\raisebox{0ex}{\mbox{$\cdot$}}%
}

\begin{document}

\empprelude{input qcg;prologues := 0;} 

\title{Quantum Computing and a Unified Approach to Fast Unitary Transforms}
\author{Sos S.~Agaian\supit{a} and Andreas Klappenecker\supit{b}\skiplinehalf
\supit{a}The City University of New York and University of Texas at San Antonio\\
\texttt{sagaian@utsa.edu}\\
\supit{b}Texas A\&M University,
Department of Computer Science,
College Station, TX 77843-3112\\
\texttt{klappi@cs.tamu.edu}
}
\maketitle
\begin{abstract}
A quantum computer directly manipulates information stored in the
state of quantum mechanical systems. The available operations have
many attractive features but also underly severe restrictions, which
complicate the design of quantum algorithms.  We present a
divide-and-conquer approach to the design of various quantum
algorithms. The class of algorithm includes many transforms which are
well-known in classical signal processing applications. We show how
fast quantum algorithms can be derived for the discrete Fourier
transform, the Walsh-Hadamard transform, the Slant transform, and the
Hartley transform.  All these algorithms use at most $O(\log^2 N)$
operations to transform a state vector of a quantum computer of
length~$N$.
\end{abstract}

\section{Introduction}
Discrete orthogonal transforms and discrete unitary transforms have
found various applications in signal, image, and video processing, in
pattern recognition, in biocomputing, and in numerous other
areas~[\citenum{agaian86,agaian88,agaian91,agaian92,ahmed75,%
andrews70,beth84,enomoto71,mali94,pratt74,selesnick99}].  Well-known examples
of such transforms include the discrete Fourier transform, the
Walsh-Hadamard transform, the trigonometric transforms such as the
Sine and Cosine transform, the Hartley transform, and the Slant
transform. All these different transforms find applications in signal
and image processing, because the great variety of signal classes
occuring in practice cannot be handeled by a single transform.

On a classical computer, the straightforward way to compute a discrete
orthogonal transform of a signal vector of length $N$ takes in general
$O(N^2)$ operations. An important aspect in many applications is to
achieve the best possible computational efficiency. The examples
mentioned above allow an evaluation with as few as $O(N\log N)$
operations or -- in the case of the wavelet transforms -- even with as
little as $O(N)$ operations. In view of the trivial lower bound of
$\Omega(N)$ operations for matrix-vector-products, we notice that
these algorithms are optimal or nearly optimal.

The rules of the game change dramatically when the ultimate limit of
computational integration is approached, that is, when information is
stored in single atoms, photons, or other quantum mechanical systems.
The operations manipulating the state of such a computer have to
follow the dictum of quantum mechanics. However, this is not
necessarily a limitation.  A striking example of the potential
speed-up of quantum computation over classical computation has been
given by Shor in 1994. He showed that integers can be factored in
polynomial time on a quantum computer. In contrast, there are no
polynomial time algorithms known for this problem on a classical
computer.

The quantum computing model does not provide a uniform speed-up for
all computational tasks. In fact, there are a number of problems which
do not allow any speed-up at all. For instance, it can be shown that a
quantum computer searching a sorted database will not have any
advantage over a classical computer. On the other hand, if we use our
classical algorithms on a quantum computer, then it will simply perform the
calculation in a similar manner to a classical computer. In order for
a quantum computer to show its superiority one needs to design new
algorithms which take advantage of quantum parallelism.

A quantum algorithm may be thought of as a discrete unitary transform
which is followed by some I/O operations.  This observation partially
explains why signal transforms play a dominant role in numerous
quantum algorithms\cite{DJ:92,Simon:94,shor94}. Another reason is that
it is often possible to find extremely efficient quantum algorithms
for the discrete orthogonal transforms mentioned above. For instance,
the discrete Fourier transform of length $N=2^n$ can be implemented
with $O(\log^2 N)$ operations on a quantum computer.

\section{Quantum Computing}
\begin{empfile}
The basic unit of information in classical computation is a bit, a
system with two distinguishable states representing logical values 0
or 1. We mentioned in the introduction that a quantum computer will
store such information in the states of a quantum mechanical system.
Suppose that the system has two distinguishable states. We will denote
these states by $\ket{0}$ and $\ket{1}$, where the notation reminds us
that these states represent the logical values 0 and 1. 

A potential candidate for the storage of a single bit is given by a
spin-$\frac{1}{2}$ particle, such as an electron, proton, or neutron.
We can choose the state with the rotation vector pointing upward
(spin-up) and the state with the rotation vector pointing downward
(spin-down) to represent 0 and 1, respectively.  However, we know from
quantum mechanics that quantum system can be in a superposition of
states. In the case of a spin-$\frac{1}{2}$ particle, a superposition
$$
\Big|\raisebox{-1.3ex}{\mbox{\epsfig{file=spin.45}}}\Big> 
= a\, \Big|\raisebox{-2ex}{\mbox{\epsfig{file=spin.0}}}\Big> 
+ b\,\Big|\raisebox{-2ex}{\mbox{\epsfig{file=spin.180}}}\Big>
$$
yields a state which rotates about a different axis. The coefficients
$a, b$ in this superposition are complex numbers, which determine this
spin axis. 

The consequent abstraction of the preceding example leads to the
notion of a quantum bit, or shortly qubit, the basic unit of information in quantum
computation. A quantum bit is given by a superposition of the states
$\ket{0}$ and $\ket{1}$ such as
$$\ket{\psi} = a \ket{0} + b\ket{1},\qquad a,b\in \C.$$ The value of a
quantum bit remains uncertain until it is measured.  A measurement
will collapse $\ket{\psi}$ to either the state $\ket{0}$ or to the
state $\ket{1}$. The coefficients $a$ and $b$ determine the probability 
of outcome of this measurement, namely 
$$ \begin{array}{c|c}
\mbox{Event} & \mbox{Probability}\\
\hline\\[-2ex]
\ket{\psi} \mbox{ collapses to } \ket{0} & |a|^2/(|a|^2+|b|^2)\\[1ex]
\ket{\psi} \mbox{ collapses to } \ket{1} & |b|^2/(|a|^2+|b|^2)
   \end{array}
$$
In either case, we will learn the outcome of the measurement.  Since
proportional states lead to the same measurement results, it is
conventially assumed that the state is normalized to length 1, i.e., it is
assumed that $|a|^2+|b|^2=1$ holds.

The measurement allows to implement a fair coin flip on a quantum
computer.  Indeed, preparing a quantum bit in the state
$\ket{\psi}=\frac{1}{\sqrt{2}}\ket{0}+\frac{1}{\sqrt{2}}\ket{1}$, and
measuring the result yields either 0 or 1. According to the above
rule, either event will occur exactly with probability 1/2.  This
example might suggest that computations on a quantum computer are
indeterministic and maybe even somewhat fuzzy. However, this is not
the case. We will see in a moment that all operations apart from
measurements are completely deterministic. The only operations that
might introduce some randomized behaviour are the measurements, which
-- as Penrose puts it -- `magnify an event from the quantum level to
the classical level'~[\citenum{penrose00}, pp.~7-8].

We discuss now the deterministic operations on a quantum computer.  We
begin with the simplest case, the operations which manipulate
the state of a single quantum bit. First of all, it should be noted that 
the states $\ket{0}$ and $\ket{1}$ can be understood as an orthonormal
basis of the complex inner product space $\C^2$. It is customary to
associate the base states $\ket{0}$
and $\ket{1}$ with the standard basis vectors 
$(1,0)^t$ and $(0,1)^t$, respectively.  
Therefore, a quantum bit in the state $a\ket{0}+b\ket{1}$
is represented by the state vector
$$ 
\left(\begin{array}{r} a\\ b\end{array} \right) = 
a\left(\begin{array}{r} 1\\ 0\end{array} \right) +
b\left(\begin{array}{r} 0\\ 1\end{array} \right).
$$ 
A deterministic operation has to realize a unitary evolution of the
quantum state, following the rules of quantum mechanics.  In other
words, a single quantum bit operation is given by a unitary operator
$U\colon \C^2\rightarrow \C^2$ acting on the state of the quantum bit.
There is a graphical notation for quantum operations. The 
schematic for such a single qubit operation $U$ is shown in Figure~\ref{fig:single}.
\begin{figure}[ht]
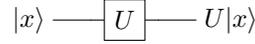
 
\vspace*{-0.5cm}
\begin{center}
\begin{emp}(50,50)
  setunit 1.5mm;
  qubits(1);
  
  labels.lft(0,btex $|x\rangle$ etex);
  wires(0.5cm);
  gate(gpos 0, btex $U$ etex);
  wires(0.5cm);
  labels.rt(0,btex $U|x\rangle$ etex);
\end{emp}
\end{center}
\caption{Schematic for a single qubit operation. The diagrams are read
from left to right, reflecting the abstraction of time
flow.}\label{fig:single}
\end{figure}

Examples of single qubit operations are given by 
$$ X = \mat{0}{1}{1}{0},\qquad Z =\mat{1}{0}{0}{-1},\qquad
H=\frac{1}{\sqrt{2}}\mat{1}{1}{1}{-1}.$$ The operation $X$ realizes a
NOT operation, $X\ket{0}=\ket{1}$ and $X\ket{1}=\ket{0}$. The
operation $Z$ implements a phase shift operation, $Z\ket{0}=\ket{0}$
and $Z\ket{1}=-\ket{1}$. An extremely useful operation is given by the
Hadamard gate~$H$, which is for instance used to create
superpositions,
$$
H\ket{0}=\frac{1}{\sqrt{2}}\ket{0}+\frac{1}{\sqrt{2}}\ket{1},\qquad
H\ket{1}=\frac{1}{\sqrt{2}}\ket{0}-\frac{1}{\sqrt{2}}\ket{1}.$$ The
Hadamard gate should be familiar to readers with a background in
signal processing or coding theory. In the following, we will keep the
notations for these gates without further notice.

The operations get more interesting in the case of multiple quantum
bits.  Quantum mechanics tells us that the state space of a combined
quantum system is given by the tensor product of the state spaces of
its parts. A remarkable consequence of this rule is that the state
space of a system with $n$ quantum bits is given by the vector space
$\C^{2^n}\cong \C^2\otimes \cdots \otimes \C^2$ ($n$-fold tensor
product). This simply means that the dimension of the state space {\em
doubles} with the addition of a single quantum bit.

The state of a system with two quantum bits can thus be described by a
vector $(a_{00},a_{01},a_{10},a_{11})^t\in \C^4$ or, isomorphically, by
the vector 
\begin{equation}\label{eq:twostate} 
\ket{\psi} = a_{00}\ket{0}\otimes\ket{0}+ a_{01}\ket{0}\otimes\ket{1}+
a_{10}\ket{1}\otimes\ket{0}+ a_{11}\ket{1}\otimes\ket{1}\in
\C^2\otimes \C^2.
\end{equation}
The latter notation is often abbreviated to $a_{00}\ket{00}+
a_{01}\ket{01}+ a_{10}\ket{10}+ a_{11}\ket{11} $.  The label $x_1x_0$
in the Dirac ket notation $\ket{x_1x_0}$ specifies a location in the
quantum memory.

A dramatic consequence of the tensor product structure of the quantum
memory can be illustrated with a single qubit operation. Suppose that
we apply a single qubit operation, say the Hadamard gate $H$, on the
least significant bit of (\ref{eq:twostate}). The resulting state is 
$$
\begin{array}{lcl}
\ket{\psi'} &=& a_{00}\ket{0}\otimes H\ket{0}+ a_{01}\ket{0}\otimes H\ket{1}+
a_{10}\ket{1}\otimes H\ket{0}+ a_{11}\ket{1}\otimes H\ket{1}\\
&=& \displaystyle\frac{1}{\sqrt{2}}\Big(
(a_{00}+a_{01})\ket{0}\otimes\ket{0} + 
(a_{00}-a_{01})\ket{0}\otimes\ket{1} + 
(a_{10}+a_{11})\ket{1}\otimes\ket{0} + 
(a_{10}-a_{11})\ket{1}\otimes\ket{1}
\Big).
\end{array}
$$ 
In more traditional mathematical notation, we can formulate this as the action of the matrix 
$$ 
\frac{1}{\sqrt{2}}
\left(
\begin{array}{rrrr}
1&1\\
1&-1\\
&&1&1\\
&&1&-1
\end{array}
\right)
\left(
\begin{array}{r}
a_{00}\\
a_{01}\\
a_{10}\\
a_{11}
\end{array}
\right)
=
\frac{1}{\sqrt{2}}\left(
\begin{array}{r}
a_{00}+a_{01}\\
a_{00}-a_{01}\\
a_{10}+a_{11}\\
a_{10}-a_{11}
\end{array}
\right).
$$ 
Therefore, the resulting operation is $(\onemat\otimes H) \ket{\psi}=
\ket{\psi'}$.  Although we act only on one quantum bit, we see every
single position of the state vector is manipulated.  This is a
striking example of quantum parallelism.  We observe that a
butterfly structure, well-known from many signal processing algorithms,
can be implemented with a single operation on a quantum computer.

The direct generalization to arbitrary single qubit operations is
shown in Figure~\ref{fig:twoqubit}. In general, a single qubit
operation is specified by a unitary $2\times 2$ matrix $U$, and the
position of the target qubit to which $U$ is applied. Suppose that the
target qubit position is $i$, then each state $\ket{x_{n-1}\dots
x_{i+1}x_ix_{i-1}\dots x_0}$ is unconditionally transformed to
$\ket{x_{n-1}\dots x_{i+1}}\otimes U\ket{x_i}\otimes \ket{x_{i-1}\dots
x_0}$, where $x_k\in \{0,1\}$.

\begin{figure}[ht]
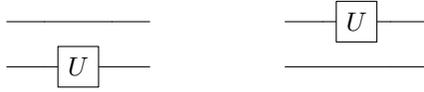

\begin{center}
\begin{emp}(50,50)
 setunit 1.5mm; 
 qubits(2);
 wires(0.5cm);
 gate(gpos 0, btex $U$ etex);
 wires(0.5cm);
 QCxcoord :=  QCxcoord+2QCstepsize;
 wires(0.5cm);
 gate(gpos 1, btex $U$ etex);
 wires(0.5cm);
\end{emp}
\vspace*{-0.5cm}
\end{center}
\caption{Single qubit operation $U$. The left side shows the schematic for 
$U$ acting on the least significant qubit; this circuit implements the
matrix $\onemat\otimes U$. The figure on the right shows the single
qubit gate acting on the most significant qubit; this circuit
implements the matrix $U\otimes\onemat$. }
\label{fig:twoqubit}
\end{figure}

We can specify more elaborate gates, which allow to create an
interaction between quantum bits. Let $C_0$ and $C_1$ be two disjoint
sets of quantum bit positions, neither of which contains the target
bit position $i$.  A conditional $U$-operation maps the state
$\ket{x_{n-1}\dots x_{i+1}x_ix_{i-1}\dots x_0}$ to the state
$\ket{x_{n-1}\dots x_{i+1}}\otimes U\ket{x_i}\otimes \ket{x_{i-1}\dots
x_0}$, in case $x_i=0$ for all $i\in C_0$ and $x_j=1$ for all $j\in
C_1$. The state remains unchanged in all other cases. The set $C_0$
describes the set of zero-conditions and $C_1$ the set of
one-conditions. In the schematics, we will use the symbol $\circ$ to
denote a zero-condition and the symbol $\bullet$ to denote a
one-condition. Figure~\ref{fig:cnot} shows the simplest, but most important, conditional quantum gate -- the controlled NOT operation. 

\begin{figure}[ht]
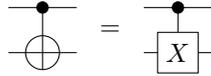

\begin{center}
\begin{emp}(50,50)
  setunit 1.5mm;
  qubits(2); 
  cnot(icnd 1, gpos 0);
  label(btex $=$ etex, (QCxcoord+1/2QCstepsize,(QCycoord[0]+QCycoord[1])/2));
  QCxcoord := QCxcoord + QCstepsize;
  gate(icnd 1, gpos 0, btex $X$ etex);
\end{emp}
\vspace*{-0.5cm}
\end{center}
\caption{The controlled NOT gate is a reversible XOR gate. The states
$\ket{00}$ and $\ket{01}$ remain unchanged, since the most significant
qubit must be 1. If the most significant bit is 1, then a NOT
operation is applied to the least significant bit. Therefore,
$\ket{10}$ is mapped to $\ket{11}$, and $\ket{11}$ is mapped to
$\ket{10}$. }\label{fig:cnot}
\end{figure}

We can use controlled NOT gates to get an interaction between
different quantum bits. For example, consider the circuit in
Figure~\ref{fig:swap}. This circuit swaps the states of the two
quantum bits.
\begin{figure}[h]
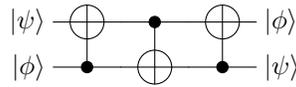

\begin{center}
\begin{emp}(50,50)
  setunit 1.5mm;
  qubits(2);
  labels.lft(0,btex $|\phi\rangle$ etex,1,btex $|\psi\rangle$ etex);
  cnot(icnd 0, gpos 1); 
  cnot(icnd 1, gpos 0); 
  cnot(icnd 0, gpos 1);
  labels.rt(0,btex $|\psi\rangle$ etex,1,btex $|\phi\rangle$ etex);
\end{emp}
\vspace*{-0.5cm}
\end{center}
\caption{Circuit which swaps the state of two quantum bits.}
\label{fig:swap}
\end{figure}

Engineering controlled $U$ operations is in general a difficult task.
We will refer to controlled NOT gates with a single control bit and to
single qubit operations as {\em elementary gates}. Elementary quantum
gates are available in all mature quantum computing technologies.  It
can be shown that it is possible to implement a general controlled $U$
operation with $O(\log N)$ elementary gates.\cite{nielson00} We will
always refer to elementary gates in gate counts, but we will use
multiply controlled $U$ gates for the sake of brevity in circuit
descriptions. There exist standard algorithms which transform these
more general gates into a sequence of elementary gates.\cite{nielson00}

\section{Divide-and-Conquer Methods}
We have seen that a number of powerful operations are available on a
quantum computer. Suppose that we want to implement a unitary or
orthogonal transform $U\in U(2^n)$ on a quantum computer. The goal
will be to find an implementation of $U$ in terms of elementary
quantum gates.  Usually, our aim will be to find first a
factorization of $U$ in terms of sparse structured unitary matrices
$U_i$,
$$ U = U_1 U_2\cdots U_k,$$ where, of course, $k$ should be small. The
philosophy being that it is often very easy to derive quantum circuits
for structured sparse matrices. For example, if we can find an
implementation with few multiply controlled unitary gates for each
factor $U_i$, then the overall circuit will be extremely efficient. 

The success of this method depends of course very much on the
availability suitable factorization of $U$. However, in the case
orthogonal transfroms used in signal processing, there are typically
numerous classical algorithms available, which provide the suitable
factorizations. It should be noted that, in principle, an exponential
number of elementary gates might be needed to implement even a
diagonal unitary matrix. Fortunately, we will see that most structured
matrices occuring in practice have very efficient implementations. In
fact, we will see that all the transforms of size $2^n\times 2^n$
discussed in the following can be implemented with merely $O(\log^2
2^n)=O(n^2)$ elementary quantum gates.

We present a simple -- but novel -- approach to derive such efficient
implementations. This approach is based on a divide-and-conquer
technique.  Assume that we want to implement a family of unitary
transforms $U_N$, where $N=2^n$ denotes the length of the
signal. Suppose further the family $U_N$ can be recursively generated
by a recursive circuit construction, for instance, such as the one
shown in Figure~\ref{fig:rec1}. We will give a generic construction
for the family of precomputation circuits Pre\_$\!$\_ and the family
of postcomputation circuits Post\_$\!$\_. This way, we obtain a fairly
economic description of the algorithms.
\begin{figure}[ht]
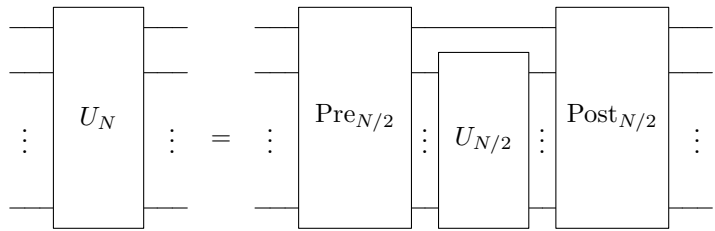

\begin{center}
\begin{emp}(50,50)
  setunit 1.5mm;
  qubits(5);
  dropwire(1,2);
  wires(2mm);  
  label(btex $\vdots$ etex,(QCxcoord,QCycoord[0]+10mm));
  wires(2mm);  
  circuit(2QCheight)(gpos 0,1,2, btex $U_N$ etex);
  wires(2mm);  
  label(btex $\vdots$ etex,(QCxcoord,QCycoord[0]+10mm));
  wires(2mm);
  label(btex $=$ etex, (QCxcoord+1/2QCstepsize,QCycoord[0]+3/2QCheight));
  QCxcoord := QCxcoord + QCstepsize;
  wires(2mm);  
  label(btex $\vdots$ etex,(QCxcoord,QCycoord[0]+10mm));
  wires(2mm);
  circuit(2.5QCheight)(gpos 0,1,2,btex $\mbox{Pre}_{N/2}$ etex);
  label(btex $\vdots$ etex,(QCxcoord,QCycoord[0]+10mm));
  circuit(2QCheight)(gpos 0,1, btex $U_{N/2}$ etex);
  label(btex $\vdots$ etex,(QCxcoord,QCycoord[0]+10mm));
  circuit(2.5QCheight)(gpos 0,1,2,btex $\mbox{Post}_{N/2}$ etex);
  wires(2mm);  
  label(btex $\vdots$ etex,(QCxcoord,QCycoord[0]+10mm));
  wires(2mm);
\end{emp}
\end{center}
\vspace*{-0.5cm}
\caption{Recursive implementation of a family of quantum circuits $U_N$. If the preparation circuit Pre$_{N/2}$ and postcomputation circuits Post$_{N/2}$ have small complexity,  then the overall circuit family will have an efficient implementation.}\label{fig:rec1}
\end{figure}

Assume that a total of $P(N)$ elementary operations are necessary to implement the precomputation circuit Pre$_{N/2}$ and the
postcomputation circuit Post$_{N/2}$. Then the overall number $T(N)$ of elementary operations can be estimated from the recurrence equation
$$ T(N) = T(N/2) + P(N).$$ 
The number of operations $T(N)$ for the recursive implementation can be estimated as follows: 

\noindent\textsc{Lemma.} 
If $P(N)\in \Theta(\log^p N)$, then $T(N)\in O(\log^{p+1} N)$.

\section{Fourier Transform}
We will illustrate the general approach by way of some examples. Our
first example is the discrete Fourier transform. A quantum algorithm
implementing this transform found a most famous application in Shor's
integer factorization algorithm.\cite{shor94} Recall that the discrete
Fourier transform $F_N$ of length $N=2^n$ can be described by the
matrix
$$ F_N = \frac{1}{\sqrt{N}}\Big(\,\omega^{jk} \Big)_{j,k=0,\dots,N-1},$$
where $\omega$ denotes a primitive $N$-th root of unity, 
$\omega=\exp(2\pi i/N)$. And $i$ denotes a square root of $-1$. 

The main observation behind the fast quantum algorithm dates at least
back to work by Danielson and Lanczos in 1942 (and is implicitly
contained in numerous earlier works). 
They noticed that the matrix
$F_N$ might be written as
$$ F_N = \frac{1}{\sqrt{2}} P_N
\cmat{F_{N/2}}{F_{N/2}}{F_{N/2}T_{N/2}}{-{F_{N/2}T_{N/2}}}$$
where $P_N$ denotes the permutation of rows given by
$P_N\ket{bx}=\ket{xb}$ with $x$ an $n-1$-bit integer, and $b$ a
single bit, and $T_{N/2}:=\diag(1,\omega,
\omega^2,\dots,\omega^{N/2-1})$ denotes the matrix of twiddle factors.

This observation allows to represent $F_N$ by the following product of
matrices:
\begin{eqnarray*}
 {F_N} & = &  P_N\cmat{F_{N/2}}{\zeromat}{\zeromat}{F_{N/2}}
\mat{\onemat_{N/2}}{\zeromat}{\zeromat}{T_{N/2}}\frac{1}{\sqrt{2}}
\mat{\onemat_{N/2}}{\onemat_{N/2}}{\onemat_{N/2}}{-\onemat_{N/2}}\\
&=&
P_N (\onemat_{2} \otimes F_{N/2}) 
                   \left(\begin{array}{cc} \onemat_{N/2} & \\ 
                          & T_{N/2} \end{array} \right)(F_2 \otimes
                   \onemat_{N/2})
\end{eqnarray*}
This factorization yields an outline of an implementation on a quantum
computer. The overall structure is shown in
Figure~\ref{fig:fourierrec}.

\begin{figure}[ht]
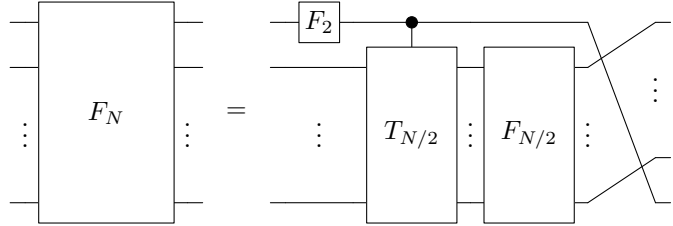

\begin{center}
\begin{emp}(50,50)
  setunit 1.5mm;
  qubits(5);
  dropwire(1,2);
  wires(2mm);
  label(btex $\vdots$ etex,(QCxcoord,QCycoord[0]+10mm));
  circuit(3QCheight)(gpos 0,1,2,btex $F_{N}$ etex);
  label(btex $\vdots$ etex,(QCxcoord,QCycoord[0]+10mm));
  wires(2mm);
  qubits(5);
  label(btex $=$ etex,(QCxcoord+QCstepsize/2,QCycoord[2]));
  dropwire(1,2);
  QCxcoord := QCxcoord + QCstepsize;
  wires(2mm);
  label(btex $\vdots$ etex,(QCxcoord+1/2QCstepsize,QCycoord[0]+10mm));
  gate(gpos 2,btex $F_2$ etex);
  circuit(2QCheight)(icnd 2, gpos 0,1,btex $T_{N/2}$ etex);
  label(btex $\vdots$ etex,(QCxcoord,QCycoord[0]+10mm));
  QCfgcolor := 0.7white;
  circuit(2QCheight)(gpos 0,1, btex $F_{N/2}$ etex);
  QCfgcolor := black;
  label(btex $\vdots$ etex,(QCxcoord,QCycoord[0]+10mm));
  qubits(5);
  dropwire(2);
  draw (QCxcoord,QCycoord[3])--(QCxcoord+QCstepsize,QCycoord[0]);
  draw (QCxcoord,QCycoord[0])--(QCxcoord+QCstepsize,QCycoord[1]);
  draw (QCxcoord,QCycoord[2])--(QCxcoord+QCstepsize,QCycoord[3]);
  dropwire(2);
  QCxcoord := QCxcoord + QCstepsize;
  wires(2mm);
  label(btex $\vdots$ etex, (QCxcoord-2mm,QCycoord[1]+10mm));
\end{emp}
\vspace*{-0.5cm}
\end{center}
\caption{The recursive structure of the quantum Fourier transform.}\label{fig:fourierrec}
\end{figure}

It remains to detail the different steps in this implementation.  The
first step is a single qubit operation, implementing a butterfly
structure. The next step is slightly more complicated. We observe that 
$T_{N/2}$ is a tensor product of diagonal matrices
$D_j=\diag(1,\omega^{2^{j-1}})$. Indeed, 
$$ T_{N/2}=D_{n-1}\otimes \dots \otimes D_2\otimes D_1.$$ Thus,
$\onemat_{N/2}\oplus T_{N/2}$ can be realized by controlled phase
shift operations, see Figure~\ref{fig:twiddle} for an example. We then
recurse to implement the Fourier transform of smaller size. The final
permutation implements the cyclic rotation of the quantum wires.
\begin{figure}[h]
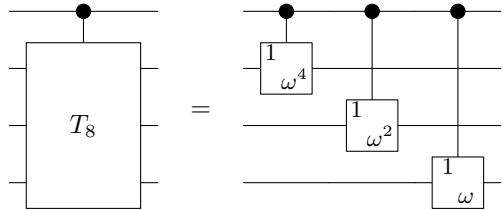

\begin{center}
\begin{emp}(50,50)
  setunit 1.9mm;
  qubits(4);
  circuit(2QCheight)(icnd 3, gpos 0,1,2, btex $T_8$ etex);
  label(btex $=$ etex, (QCxcoord + 1/2QCstepsize, QCycoord[1]+1/4QCheight));
  QCxcoord := QCxcoord + QCstepsize;
  gate(icnd 3, gpos 2, btex \small $\begin{array}{r@{\,}l} 1\\ 
                        & \omega^4\end{array}$ etex); 

  gate(icnd 3, gpos 1, btex \small $\begin{array}{r@{\,}l} 1\\ 
                        & \omega^2\end{array}$ etex);
  gate(icnd 3, gpos 0, btex \small $\begin{array}{r@{\,}l} 1\\ 
                        & \omega\end{array}$ etex);

\end{emp}
\vspace*{-0.5cm}
\end{center}
\caption{Implementation of the twiddle matrix $\onemat_8\oplus T_{8}$.}
\label{fig:twiddle}
\end{figure}

The complexity of the quantum Fourier transform can be
estimated as follows. 
If we denote by $R(N)$ the number of gates necessary to implement the
DFT of length $N=2^n$ on a quantum computer, then Figure~\ref{fig:fourierrec}
implies the recurrence relation
$$ R(N)=R(N/2)+\Theta(\log N)$$
which leads to the estimate $R(N)=O(\log^2N)$.

It should be noted that all permutations 
$ P_N(\onemat_2\otimes P_{N/2})\dots
(\onemat_{N-2}\otimes P_4)$ at the end
can be combined into a single permutation of quantum wires. 
The resulting permutation
is the bit reversal, see Figure~\ref{bitreversal}.
\begin{figure}[ht]
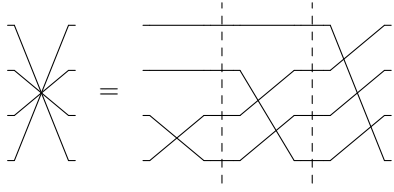

\begin{center}
\begin{emp}(50,50)
  setunit 1.5mm;
  qubits(4);
  perm(cyc 0,3, cyc 1,2);
  label(btex $=$ etex, (QCxcoord+1/2QCstepsize, QCycoord[1]+1/2QCheight));
  QCxcoord := QCxcoord + QCstepsize;
  perm(cyc 0,1);
  wires(1.5mm);
  draw (QCxcoord,QCycoord[0]-1/2QCheight)--(QCxcoord,QCycoord[3]+1/2QCheight) dashed evenly;
  wires(1.5mm);
  perm(cyc 2,0,1);
  wires(1.5mm);
  draw (QCxcoord,QCycoord[0]-1/2QCheight)--(QCxcoord,QCycoord[3]+1/2QCheight) dashed evenly;
  wires(1.5mm);
  perm(cyc 3,0,1,2);
\end{emp}
\vspace*{-0.5cm}
\end{center}
\caption{\label{bitreversal} 
The bit reversal permutation resulting from $P_8(\onemat_2\otimes P_4)(\onemat_4\otimes P_2)$.} 
\end{figure}

\noindent\textsc{Remark.} Another explanation of the discrete Fourier
transform algorithm is contained in [\citenum{klappieff}]. Note that
the row permutations are mistaken in that article. An approximate
version of the discrete Fourier transform has been proposed by
Coppersmith\cite{coppersmith94}, which saves some operations.

\section{The Walsh-Hadamard Transform} 
The Walsh-Hadamard transform $W_N$ is maybe the simplest instance of the
recursive approach. This transform is defined by the Hadamard gates $W_2=H$ in the case of signals of length 2. For signals of larger length, the transform is defined by 
$$ W_N = (\onemat_2\otimes W_{N/2})(H\otimes \onemat_{N/2}).$$ 
This yields the recursive implementation shown in Figure~\ref{fig:walsh}.
\begin{figure}[ht]
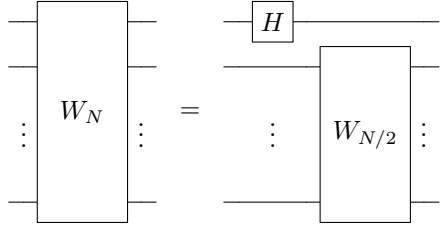

\begin{center}
\begin{emp}(50,50)
  setunit 1.5mm;
  qubits(5);
  dropwire(1,2);
  wires(2mm);
  label(btex $\vdots$ etex,(QCxcoord,QCycoord[0]+10mm));
  circuit(2QCheight)(gpos 0,1,2,btex $W_N$ etex);
  label(btex $\vdots$ etex,(QCxcoord,QCycoord[0]+10mm));
  wires(2mm);

  label(btex $=$ etex, (QCxcoord + 1/2QCstepsize, QCycoord[0]+2QCheight));
  QCxcoord := QCxcoord + QCstepsize;
  wires(2mm);
  label(btex $\vdots$ etex,(QCxcoord+1/2QCstepsize,QCycoord[0]+10mm));
  gate(gpos 2, btex $H$ etex);
  circuit(2QCheight)(gpos 0,1, btex $W_{N/2}$ etex);
  label(btex $\vdots$ etex,(QCxcoord,QCycoord[0]+10mm));
  wires(2mm);
\end{emp}
\end{center}
\caption{Recursive implementation of the Walsh-Hadamard transform.} \label{fig:walsh}
\end{figure}
   
Since $P(N)=\Theta(1)$, the Lemma in Section~3 shows that the number
of operations $T(N)\in O(\log N)$. It is of course trivial to see that in this case exactly $\log N$ operations are needed.

\section{The Slant Transform}
The Slant transform is used in image processing for the representation
of images with many constant or uniformly changing gray levels.  The
transform has good energy compaction properties. It is used in Intel's
`Indeo' video compression and in numerous still image compression
algorithms.

The Slant transform $S_N$ is defined for signals of length $N=2$ by the Hadamard matrix 
$$ S_2 = H = \frac{1}{\sqrt{2}}\mat{1}{1}{1}{-1},$$
and for signals of length $N=2^k$, $N>2$, by 
\begin{equation} \label{eq:slantrec}
S_N = 
Q_N\mat{S_{N/2}}{\mathbf{0}_{N/2}}{\mathbf{0}_{N/2}}{S_{N/2}},
\end{equation}
where
$\mathbf{0}_{N/2}$ denotes the all-zero matrix, and $Q_N$ is given by the matrix product  
\begin{equation}\label{eq:slant}
 Q_N = P_N^a( \onemat_{N/2} \oplus \widehat{Q}_{N} ) (H\otimes \onemat_{N/2}) P_{N}^b.
\end{equation}
The matrices in (\ref{eq:slant}) are defined as follows (see also~[\citenum{yang97}]):
$\onemat_{N/2}$ is the identity matrix, $H$ is the Hadamard matrix, and
$P_N^a$ realizes the transposition $(1,N/2)$, that is, 
$$P_N^a\ket{1}=\ket{N/2},\quad  P_N^a\ket{N/2}=\ket{1},\quad \mbox{and} 
\quad P_N^a\ket{x}=\ket{x}\; \mbox{otherwise.}$$
The matrix $P_N^b$ is defined by $P_N^b\ket{x}=\ket{x}$ for all $x$ except in the
case $x= N/2+1$, where it yields the phase change
$P_N^b\ket{N/2+1}=-\ket{N/2+1}$. Finally
$$ \widehat{Q}_{N} = 
\left(\begin{array}{cc}
A_N 
 & \zeromat\\
\zeromat & \onemat_{(\frac{N}{2}-2)}
      \end{array}
\right),
\qquad A_N = \left(
\begin{array}{cc}
a_N & b_N\\
-b_N & a_N
\end{array}
\right),
$$ 
where $a_N$ and $b_N$ are recursively defined by $a_2=1$ and 
$$ b_N = \frac{1}{\sqrt{1+4(a_{N/2})^2}}
\qquad\mbox{and}\qquad a_N = 2b_Na_{N/2}.$$ 
It is easy to check that $A_N$ is a unitary matrix. 

The definition of the Slant transform suggests the following
implementation.  Equation (\ref{eq:slantrec}) tells us that the input
signal of a Slant transform of length $N$ is first processed by two
Slant transforms of size $N/2$, followed by a circuit implementing $Q_N$. 
We can write equation (\ref{eq:slantrec}) in the form 
$$ S_N = 
Q_N\mat{S_{N/2}}{\mathbf{0}_{N/2}}{\mathbf{0}_{N/2}}{S_{N/2}} = 
Q_N(\onemat_2 \otimes S_{N/2}).
$$ 
The tensor product structure $\onemat_2 \otimes S_{N/2}$ is compatible
with our decomposition into quantum bits. This means that a single
copy of the circuit $S_{N/2}$ acting on the lower significant bits
will realize this part. It remains to give an implementation for $Q_N$.
Equation~(\ref{eq:slant}) describes $Q_N$ as a product of four sparse
matrices, which are easy to implement.  Indeed, the matrix $P^b_N$ is
realized by conditionally excerting the phase gate $Z$. The matrix
$H\otimes \onemat_{N/2}$ is implemented by a Hadamard gate $H$ acting
on the most significant bit. A conditional application of $A_N$
implements the matrix $\onemat_{N/2}\oplus \widehat{Q}_N$. A
conditional swap of the least and the most significant qubit realizes
$P_N^a$, that is, three multiply controlled NOT gates implement $P_N^a$. 
The quantum circuit realizing this implementation is depicted in Figure~\ref{fig:slant}. 
\begin{figure}[htb]
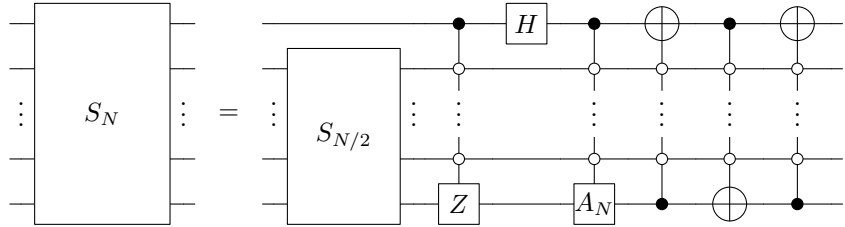

\begin{center}
\begin{emp}(50,50)
  setunit 3mm;
  qubits(5);
  dropwire(2);

  wires(unit); 
  label(btex $\vdots$ etex, (QCxcoord,QCycoord[1]+14mm));
  circuit(3*QCheight)(gpos 0,1,2,3,btex $S_N$ etex);
  label(btex $\vdots$ etex, (QCxcoord,QCycoord[1]+14mm));
  wires(unit);
  qubits(5);
  label(btex $=$ etex,(QCxcoord+1/2QCstepsize,QCycoord[2]));
  QCxcoord := QCxcoord + QCstepsize;
  dropwire(2);
  wires(unit);
  label(btex $\vdots$ etex, (QCxcoord,QCycoord[1]+14mm));
  circuit(2.5*QCheight)(gpos 0,1,2,btex $S_{N/2}$ etex);
  label(btex $\vdots$ etex, (QCxcoord,QCycoord[1]+14mm));
  wires(unit);
  xsave := QCxcoord;
  gate(icnd 3, ocnd 1,2, gpos 0,btex $Z$ etex);
  gate(gpos 3, btex $H$ etex);
  gate(icnd 3, ocnd 1,2, gpos 0, btex $A_N$ etex);
  gate(icnd 0, ocnd 1,2, gpos 3, "not");
  gate(icnd 3, ocnd 1,2, gpos 0, "not");
  gate(icnd 0, ocnd 1,2, gpos 3, "not");
  wires(unit);
  
  qubits(5);
  path rec;
  rec := (xsave, QCycoord[1]+2unit)--(xsave+6QCstepsize,QCycoord[1]+2unit)--(xsave+6QCstepsize,QCycoord[3]-2unit)--(xsave,QCycoord[3]-2unit)--cycle;
  fill rec withcolor white;
  draw rec withcolor white;
  label(btex $\vdots$ etex,(xsave+1/2QCstepsize,QCycoord[1]+14mm));
  label(btex $\vdots$ etex,(xsave+5/2QCstepsize,QCycoord[1]+14mm));
  label(btex $\vdots$ etex,(xsave+7/2QCstepsize,QCycoord[1]+14mm));
  label(btex $\vdots$ etex,(xsave+9/2QCstepsize,QCycoord[1]+14mm));
  label(btex $\vdots$ etex,(xsave+11/2QCstepsize,QCycoord[1]+14mm));

\end{emp}
\vspace*{-0.5cm}
\end{center}
\caption{Implementation of the Slant transform. The
recursive step is realized by a single Slant transform of size
$S_{N/2}$. The next three gates
implement $P_N^b$, $H\otimes \onemat_{N/2}$, and $\onemat_{N/2}\oplus
\widehat{Q}_N$, respectively. The last three gates implement
$P_N^a$. Thus, the implementation of $Q_N$ totals five multiply controlled gates and one single qubit gate.
}
\label{fig:slant}
\end{figure}

\begin{theorem} 
The Slant transform of length $N=2^k$ can be realized on a quantum
computer with at most $O(\log^2 N)$ elementary operations (that is, controlled NOT gates and single qubit gates), assuming
that additional workbits are available.
\end{theorem}
\proof Recall that a multiply controlled gate can be
expressed with at most $O(\log N)$ elementary operations 
as long as
additional workbits are available.\cite{nielson00} 
It follows from the Lemma in Section~3 that at most $O(\log^2 N)$ elementary operations are needed to implement the Slant transform.~\qed

\section{The Hartley Transform}
The discrete Hartley transform $H_N$ is defined for signals of length
$N=2^n$ by the matrix
$$ H_N = \frac{1}{\sqrt{N}}\Big(\cos(2\pi k\ell)+\sin(2\pi
k\ell)\Big)_{k,\ell=0,\dots,N-1}.$$ The discrete Hartley transform is
very popular in classical signal processing, since it requires only
real arithmetic but has similar properties. In particular, there are
classical algorithms available, which outperform the fastest Fourier
transform algorithms. We derive a fast quantum algorithm for this
transform, again based on a recursive divide-and-conquer algorithm.  A
fast algorithm for the discrete Hartley transform based on a
completely different approach has been discussed by Klappenecker and
R\"otteler\cite{klappieff}.

The Hartley transform can be recursively represented as\cite{wickerhauser93} 
\begin{equation}\label{eq:hartley} H_N = \frac{1}{\sqrt{2}} 
\mat{\onemat_{N/2}}{\onemat_{N/2}}{\onemat_{N/2}}{-\onemat_{N/2}}
\cmat{\onemat}{}{}{BC_{N/2}}
\mat{H_{N/2}}{}{}{H_{N/2}} Q_N
\end{equation}
where $Q_N$ is the permutation $Q_N\ket{xb}=\ket{bx}$, with $b$ a single bit, separating the even indexed samples and the odd indexed samples; for instance, $Q_8(x_0,x_1,x_2,x_3,x_4,x_5,x_6,x_7)^t=(x_0,x_2,x_4,x_6,x_1,x_3,x_5,x_7)^t$.  
The matrix $BC_{N/2}$ is given by 
$$
BC_{N/2}= \cmat{1}{}{}{CS_{N/2-1}}, \quad\mbox{with}\quad
\left(\begin{array}{ccccccc}
c_N^1&&&&&&s_N^1\\
& \bdots &&&&\adots\\
&&c_N^{N/4-1}&&s_N^{N/4-1}\\
&&&1\\
&&s_N^{N/4-1}&&-c_N^{N/4-1}\\
&\adots &&&& \bdots\\
s_N^1&&&&&&-c_N^1
      \end{array}
\right)
$$ 
The equation (\ref{eq:hartley}) leads to the implementation sketched in Figure~\ref{fig:hartley}.
\begin{figure}[ht]
\begin{center}
\begin{emp}(50,50)
  
  setunit 1.5 mm;
  qubits(5);
  dropwire(1,2);
  wires(2mm);
  label(btex $\vdots$ etex,(QCxcoord,QCycoord[0]+10mm));
  circuit(2QCheight)(gpos 0,1,2, btex $H_{N}$ etex);
  label(btex $\vdots$ etex,(QCxcoord,QCycoord[0]+10mm));
  wires(2mm);
  label(btex $=$ etex, (QCxcoord+1/2QCstepsize,QCycoord[0]+2QCheight));
  QCxcoord := QCxcoord + QCstepsize;
 
  qubits(5);
  dropwire(2,3);
  wires(2mm);
  label(btex $\vdots$ etex,(QCxcoord,QCycoord[1]+10mm));
  draw (QCxcoord, QCycoord[0])--(QCxcoord+QCstepsize,QCycoord[2]);
  draw (QCxcoord, QCycoord[0]+QCheight)--(QCxcoord+QCstepsize,QCycoord[0]);
  draw (QCxcoord, QCycoord[2])--(QCxcoord+QCstepsize,QCycoord[2]-QCheight);
  QCxcoord := QCxcoord + QCstepsize;
  qubits(5);
  dropwire(1,2);
  wires(2mm);
  label(btex $\vdots$ etex,(QCxcoord,QCycoord[0]+10mm));
  circuit(2QCheight)(gpos 0,1, btex $H_{N/2}$ etex);
  label(btex $\vdots$ etex,(QCxcoord,QCycoord[0]+10mm));
  circuit(2QCheight)(icnd 2, gpos 0,1, btex $BC_{N/2}$ etex);
  label(btex $\vdots$ etex,(QCxcoord,QCycoord[0]+10mm));
  gate(gpos 2,btex $H$ etex);
  label(btex $\vdots$ etex,(QCxcoord,QCycoord[0]+10mm));
  wires(2mm);
\end{emp}
\end{center}
\caption{Recursive implementation of the Hartley transform.}
\label{fig:hartley}
\end{figure}

It remains to describe the implementation of $BC_{N/2}$.  It will be
instructive to detail the action of the matrix $BC_{N/2}$ on a state
vector of $n-1$ qubits.  We will need a few notations first.  Denote
by $\ket{bx}$ a state vector of $n-1$ qubits, where $b$ denotes a
single bit and $x$ an $n-2$ bit integer. We denote by $x'$ the two's
complement of $x$.  We mean by $x=\mathbf{0}$ the number $0$ and by
$\mathbf{1}$ the number $2^{n-2}-1$, that is, $\mathbf{1}$ has all
bits set and $\mathbf{0}$ has no bit set. Then the action of
$BC_{N/2}$ on $\ket{bx}$ is given by
$$
\begin{array}{ll}
BC_{N/2}\ket{0\mathbf{0}} = \ket{0\mathbf{0}},\quad &
BC_{N/2}\ket{0 y} = c_N^y\ket{0 y}+ s_N^y\ket{1 y'},\\ 
BC_{N/2}\ket{0\mathbf{1}} = \ket{0\mathbf{1}},&
BC_{N/2}\ket{1 y} = s_N^{y'}\ket{0 y'}-c_N^{y'}\ket{1 y},
\end{array}
$$ 
where $s_N^k=\sin(2\pi k/N)$ and $c_N^k=\cos(2\pi k/N)$. 
\begin{figure}[ht]
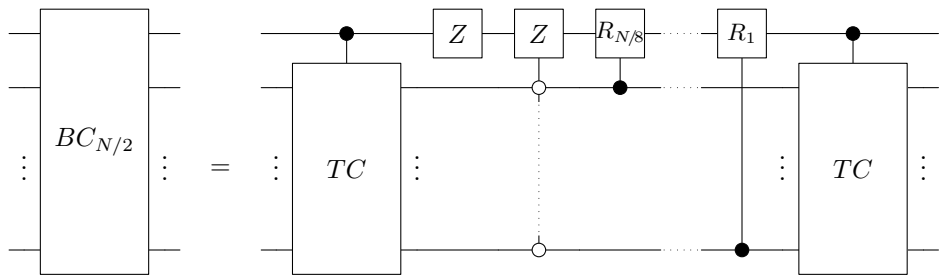

\begin{center}
\begin{emp}(50,50)
  setunit 1.8mm;
  qubits(5);
  dropwire(1,2);
  wires(2mm);
  label(btex $\vdots$ etex,(QCxcoord,QCycoord[0]+12mm));
  circuit(2QCheight)(gpos 0,1,2, btex $BC_{N/2}$ etex);
  label(btex $\vdots$ etex,(QCxcoord,QCycoord[0]+12mm));
  wires(2mm);
  label(btex $=$ etex, (QCxcoord+1/2QCstepsize, QCycoord[0]+1.5QCheight));
  QCxcoord := QCxcoord + QCstepsize;
  
  wires(2mm);
  label(btex $\vdots$ etex,(QCxcoord,QCycoord[0]+12mm));
  circuit(2QCheight)(icnd 2, gpos 0,1, btex $TC$ etex);
  label(btex $\vdots$ etex,(QCxcoord,QCycoord[0]+12mm));
  gate(gpos 2,btex $Z$ etex);
  gate(ocnd 0,1, gpos 2, btex $Z$ etex);

  path rec;
  rec := (QCxcoord-2/3QCstepsize,QCycoord[0]+2mm)
     --(QCxcoord-1/3QCstepsize,QCycoord[0]+2mm)
     --(QCxcoord-1/3QCstepsize,QCycoord[1]-2mm)
     --(QCxcoord-2/3QCstepsize,QCycoord[1]-2mm)--cycle;
  fill rec withcolor white;
  draw rec withcolor white;
  draw (QCxcoord-1/2QCstepsize,QCycoord[0]+2mm)--(QCxcoord-1/2QCstepsize,QCycoord[1]-2mm) dashed withdots scaled 1/2;
  gate(icnd 1, gpos 2, btex \small $R_{N\!/\!8}$ etex);
  draw (QCxcoord,QCycoord[0])
     --(QCxcoord+1/2QCstepsize, QCycoord[0]) dashed withdots scaled 1/2;
  draw (QCxcoord,QCycoord[1])
     --(QCxcoord+1/2QCstepsize, QCycoord[1]) dashed withdots scaled 1/2;
  draw (QCxcoord,QCycoord[2])
     --(QCxcoord+1/2QCstepsize, QCycoord[2]) dashed withdots scaled 1/2;
  QCxcoord := QCxcoord + 1/2QCstepsize;
  gate(icnd 0, gpos 2, btex \small $R_{1}$ etex);
  label(btex $\vdots$ etex,(QCxcoord,QCycoord[0]+12mm));
  circuit(2QCheight)(icnd 2, gpos 0,1, btex $TC$ etex);
  label(btex $\vdots$ etex,(QCxcoord,QCycoord[0]+12mm));
  wires(2mm);
\end{emp}
\end{center}
\caption{Implementation of the matrix $BC_{N/2}$.} 
\label{fig:BC}
\end{figure}

We are now in the position to describe the implementation of
$BC_{N/2}$ shown in Figure~\ref{fig:BC}.  In the first step, the least
$n-2$ qubits are conditionally mapped to their two's complement. More
precisely, the input signal $\ket{bx}$ is mapped to $\ket{bx'}$ if
$b=1$, and does not change otherwise.  Thus, the circuit $TC$
implements the involutary permutation corresponding to the two's
complement operation. This can be done with $O(n)$ elementary gates,
provided that sufficient workspace is available.\cite{vedral96} In the
next step, a sign change is done if $b=1$, that is, $\ket{1x}\mapsto
-\ket{1x}$, unless the input $x$ was equal to zero,
$\ket{1\mathbf{0}}\mapsto\ket{1\mathbf{0}}$. The next step is a
conditioned cascade of rotations. The least significant bits determine
the angle of the rotation on the ($n-1$st) most significant qubit. The $k$th qubits exerts a rotation,
$$ R_{2^k} = \mat{\cos(2\pi 2^k/N)}{-\sin(2\pi 2^k/N)}{\sin(2\pi 2^k/N)}{\cos(2\pi 2^k/N)},
$$
on the most significant qubit. Finally, another two's
complement circuit is conditionally applied to the state.

One readily checks that the implementation indeed maps
$BC_{N/2}\ket{0\mathbf{0}}$ to $\ket{0\mathbf{0}}$ and
$BC_{N/2}\ket{1\mathbf{0}}$ to $\ket{1\mathbf{0}}$. The input
$\ket{0x}$ is mapped to $c_N^{x}\ket{0x}+s_N^{x}\ket{1x'}$, as
desired. Assume that the input is $\ket{1x}$ with $x\neq 0$. Then the
state is changed to $\ket{1x'}$ by the circuit $TC$, and after that
its sign is changed, which yields $-\ket{1x'}$. The rotations map this state to $s_N^{x'}\ket{0x'}-c_N^{x'}\ket{1x'}$. The final conditional two's complement operation yields the state $s_N^{x'}\ket{0x'}-c_N^{x'}\ket{1x}$, which is exactly what we want. 

The inital permutation, the circuit $BC_{N/2}$ and the Hadamard gate
in Figure~\ref{fig:hartley} can be implemented with $\Theta(\log N)$
elementary gates. It is crucial that additional workbits are available, otherwise the complexity will increase to $\Theta(\log^2 N).$ The Lemma in Section~3 then completes the proof of the following theorem:
\begin{theorem}
There exists a recursive implementation of the discrete Hadamard
transform $H_N$ on a quantum computer with $O(\log^2 N)$ elementary
gates (that is, controlled NOT gates and single qubit gates), assuming
that additional workbits are available.
\end{theorem}

\end{empfile}
\section{Conclusions}
We have presented a new approach to the design of quantum
algorithms. The method takes advantage of an divide-and-conquer
approach. We have illustrated the method in the design of quantum
algorithms for the Fourier, Walsh, Slant, and Hartley transforms.  The
same method can be applied to derive fast algorithms for various
discrete Cosine transforms.  It might seem surprising that
divide-and-conquer methods have not been previously suggested in
quantum computing (to the best of our knowledge). One reason might be
that the quantum circuit model implements only straight-line programs.
We defined recursions on top of that model, similar to macro
expansions in many classical programming languages. The benefit is
that many circuits can be specified in a very lucid way.

It should be emphasized that our divide-and-conquer approach is
completely general. It can be applied to a much larger class of
circuits, and is of course not restricted to signal processing
applications. Moreover, it should be emphasized that many variations
of this method are possible.  We would like to encourage the reader to
work out a few examples -- quite often this is a simple exercise.


\begin{thebibliography}{10}
\bibitem{agaian86}
S.~Agaian, ``Optimal algorithms of fast orthogonal transforms and their
  implementation on computers,'' {\em Kibernetika I Vichislitelnaya Tekhnika}
  {\bf 2}, pp.~231--319, 1986.
\newblock (In Russian).

\bibitem{agaian88}
S.~Agaian and D.~Gevorkian, ``Complexiy and parallel algorithms of discrete
  orthogonal transforms,'' {\em Kibernetika I Vichislitelnaya Tekhnika} {\bf
  4}, pp.~124--169, 1988.
\newblock (In Russian).

\bibitem{agaian91}
S.~Agaian and V.~Duvalyan, ``On {S}lant transforms,'' {\em Pattern Recognition
  and Image Analysis} {\bf 1}(3), pp.~317--326, 1991.

\bibitem{agaian92}
S.~Agaian and D.~Gevorkian, ``Synthesis of a class of orthogonal transforms --
  {P}arallel {SIMD}-algorithms and specialized processors,'' {\em Pattern
  Recognition and Image Analysis} {\bf 2}(4), pp.~394--408, 1992.

\bibitem{ahmed75}
N.~Ahmed and K.~Rao, {\em Orthogonal Transforms for Digital Signal Processing},
  Springer-Verlag, New York, 1975.

\bibitem{andrews70}
H.~Andrews and K.~Caspary, ``A generalized technique for spectral analysis,''
  {\em IEEE Trans. Computers} {\bf 19}, pp.~16--25, 1970.

\bibitem{beth84}
T.~Beth, {\em Verfahren der schnellen {F}ourier-{T}ransformation}, Teubner,
  Stuttgart, 1984.
\newblock (In German).

\bibitem{enomoto71}
H.~Enomoto and K.~Shibata, ``Orthogonal transform system for television
  signals,'' {\em IEEE Trans. Electromagn. Compat.} {\bf 13}, pp.~11--17, 1971.

\bibitem{mali94}
P.~Mali and D.~Duta~Majumder, ``An analytical comparative study of a class of
  discrete linear basis transforms,'' {\em IEEE Trans. on Systems, Man, and
  Cybernetics} {\bf 24}(3), pp.~531--535, 1994.

\bibitem{pratt74}
W.~Pratt, L.~Welch, and W.~Chen, ``Slant transform for image coding,'' {\em
  IEEE Trans. Commun.} {\bf 22}, pp.~1075--1093, 1974.

\bibitem{selesnick99}
I.~Selesnick, ``The slantlet transform,'' {\em IEEE Trans. on Signal
  Processing} {\bf 47}(5), pp.~1304--1313, 1999.

\bibitem{DJ:92}
D.~Deutsch and R.~Jozsa, ``Rapid solution of problems by quantum computation,''
  {\em Proc. R. Soc. Lond. A} {\bf 439}, pp.~553--558, 1992.

\bibitem{Simon:94}
D.~R. Simon, ``On the power of quantum computation,'' in {\em Proceedings of
  the 35th Annual Symposium on Foundations of Computer Science},  pp.~116--123,
  IEEE Computer Society Press, (Los Alamitos, CA), 1994.

\bibitem{shor94}
P.~W. Shor, ``{Algorithms for Quantum Computation: Discrete Logarithm and
  Factoring},'' in {\em Proc. FOCS 94},  pp.~124--134, IEEE Computer Society
  Press, 1994.

\bibitem{penrose00}
R.~Penrose~et al., {\em The Large, the Small and the Human Mind}, Canto
  edition, Cambridge University Press, Cambridge, 2000.

\bibitem{nielson00}
M.~Nielson and I.~Chuang, {\em Quantum Computing and Quantum Information},
  Cambridge University Press, 2000.

\bibitem{klappieff}
A.~Klappenecker and M.~R{\"o}tteler, ``On the irresistible efficieny of signal
  processing methods in quantum computing,'' in {\em Proceedings of First
  International Workshop on Spectral Techniques and Logic Design for Future
  Digital Systems, Tampere, Finland, June 2-3, 2000},  R.~Creutzburg and
  K.~Egiazarian, eds., {\em TICSP} {\bf 10}, pp.~483--497, TTKK, (Monistamo),
  2000.

\bibitem{coppersmith94}
D.~Coppersmith, ``An approximate {F}ourier transform useful in quantum
  factoring,'' Research Report RC 19642, IBM, 19994.

\bibitem{yang97}
J.-F. Yang and C.-P. Fan, ``Centralized fast slant transform algorithms,'' {\em
  IEICE Trans. Fundamentals} {\bf E80-A}(4), pp.~705--711, 1997.

\bibitem{wickerhauser93}
V.~Wickerhauser, {\em Adapted Wavelet Analysis from Theory to Software}, A.K.
  Peters, Wellesley, 1993.

\bibitem{vedral96}
V.~Vedral, A.~Barenco, and A.~Ekert, ``Quantum networks for elementary
  arithmetic operations,'' {\em Phys.~Rev.~A} {\bf 54}, pp.~147--153, 1996.

\end{thebibliography}
\end{document}